\newcommand\has{\hat{S}}
\newcommand\deq{\delta Q_{rev}}
\newcommand\pa{\partial}

\newcommand\beq{\begin{equation}}
\newcommand\eeq{\end{equation}}
\newcommand\beqn{$$}
\newcommand\eeqn{$$}
\newcommand\beqnl{\begin{eqnarray}}
\newcommand\beqna{\begin{eqnarray*}}
\newcommand\eeqna{\end{eqnarray*}}
\newcommand\eeqnl{\end{eqnarray}}

 \def\NN{\hbox{\sf I\kern-.13em\hbox{N}}}
 \def\HH{\hbox{\sf I\kern-.13em\hbox{H}}}
 \def\DD{\hbox{\sf I\kern-.13em\hbox{D}}}
 \def\RR{\hbox{\sf I\kern-.14em\hbox{R}}}
 \def\CC{\hbox{\sf I\kern-.44em\hbox{C}}}
 \def\ZZ{{\hbox{\sf Z\kern-.43emZ}}}
 \def\QQ{\hbox{\sf C\kern -.48emQ}}
 \def\Cc{\hbox{\sf C\kern -.47em {\raise .48ex \hbox{$\scriptscriptstyle |$}}
   \kern-.5em {\raise .48ex \hbox{$\scriptscriptstyle |$}} }}
 \def\Qq{\hbox{\sf Q\kern -.57em {\raise .48ex \hbox{$\scriptscriptstyle |$}}
   \kern-.55em {\raise .48ex \hbox{$\scriptscriptstyle |$}} }}

\documentstyle[prl,aps,multicol,psfig]{revtex}
\begin{document}
\draft

\vskip1pc

\title{Homogeneity as a bridge between Carath\'eodory and Gibbs}
\author{F. Belgiorno\footnote{E-mail address: belgiorno@mi.infn.it}}
\address{Dipartimento di Fisica, Universit\`a degli Studi di Milano, 
Via Celoria 16, 20133 
Milano, Italy}
\date{\today}
\maketitle

\begin{abstract}

In this paper we show that the homogeneity of the Pfaffian form 
representing the infinitesimal heat exchanged reversibly $\deq$ 
by a thermodynamic 
system allows to find immediately and explicitly an integrating factor. 
An interesting bridge between Carath\'eodory's approach to thermodynamics 
and Gibbs' approach is established.

\end{abstract}


\section{Introduction}

One can distinguish in general between two main formal approaches to 
thermodynamics.  
On one side there is the approach due to  
Carath\'eodory, which is based on the integrability of the 
Pfaffian form $\deq$ 
\cite{caratheodory,chandrasekhar,buchdahl,beattie,bazarov,wilson,landsberg,kestin,bernstein,boyling,fleming,frankel} and represents the development 
of the line of thought 
which starts with Clausius and Kelvin. On the other side one finds  
the approach due to Gibbs, 
in which $S$ is postulated to be an extensive concave function 
of the extensive variables \cite{gibbsbook,callen,tisza}; 
further developments 
of this approach, from the point of view of its axiomatization, 
find their latest realization in Ref. \cite{liebyng}.\\ 
Our aim in this paper is not to propose an axiomatics for 
thermodynamics from first principles, but simply to discuss a 
formalism which represents 
a straightforward link between Carath\'eodory's approach and 
Gibbs' approach to thermodynamics and to propose a set of constructive 
assumptions which lead to standard thermodynamics.\\
We know that Carath\'eodory's postulate of adiabatic inaccessibility  
ensures the integrability of the Pfaffian form $\deq$  
\cite{buchdahl,beattie,bazarov,wilson,landsberg,bernstein,boyling};  
in particular the so-called metrical entropy $S$ and 
the so-called absolute temperature $T$ exist and 
\beqn
\delta Q_{rev}=T\; dS.
\eeqn
We choose to work in the framework of  Carath\'eodory's approach, 
thus we assume the integrability of the Pfaffian form 
$\deq$. Moreover, we apply this formalism to (maybe multi-component) 
homogeneous simple systems, 
and choose the extensive variables $(U,V,X^1,\ldots,X^n)$ as independent 
coordinates in the equilibrium thermodynamics space 
[for ``extensive'' we mean 
``positively homogeneous functions of degree one'']. 
Notice that these variables 
characterize Gibbs' approach in the entropy representation, 
they   characterize the so-called Gibbs space \cite{tisza}; 
moreover they can be considered as the natural variables 
to be concerned with in the framework of an 
approach where the entropy and the 
temperature have not yet been introduced. In the following, 
the main assumptions of our construction are indicated as 
h1),h2),h3),h4),h5),h6),h7),h8),h9),h10) in the text. We start from h1):\\ 
\\
{\bf h1}) 
{\sf In our approach the homogeneity of the system is translated into the 
homogeneity of degree one of the integrable Pfaffian form} 
\beq
\delta Q_{rev}=dU+p\; dV-\sum_{i}\; \xi_i\; dX^i.
\label{heat}
\eeq
This means that, under the rescaling 
$(U,V,X^1,\ldots,X^n)\mapsto (\lambda U,\lambda V,
\lambda X^1,\ldots,\lambda X^n)$ one find 
$\deq \mapsto \lambda\; \deq$. See the following section for 
more details. $p,\xi_1,\ldots,\xi_n$ are intensive variables.\\ 
\\
From a physical point of view, the homogeneity of $\delta Q_{rev}$ 
can be phenomenologically inferred by the observation of 
the extensivity of $U,V,X^1,\ldots,X^n$ and 
the intensivity of $p,\xi_1,\ldots,\xi_n$ 
in the case of standard thermodynamic systems.  
Our main point is the following. 
Because of a symmetry of the Pfaffian form 
for homogeneous systems, we are able to find an explicit 
integrating factor for $\deq$ and to construct a potential 
which is identified with the entropy, in particular, with 
the entropic fundamental equation of Gibbs. In particular, 
both the entropy and 
the temperature appear as derived quantities.  
More general constructions for thermodynamics under the hypothesis 
of a non-trivial symmetry for $\deq$ are studied in \cite{qobh,belsym}.

\section{Homogeneous Pfaffian forms}

In the following, we assume that the  
differential 1-form $\omega$ is at least $C^1$ in the domain 
${\cal D}$. We define $\Omega^k ({\cal D})$ as the set of all the k-forms 
(with a suitable degree of smoothness) defined on ${\cal D}$. 
We give below some relevant 
definitions. 
\\
By ``symmetry'' for a  
differential 1-form $\omega$ we mean that, 
if $Z$ is a vector field and 
$L_Z: \Omega^1 ({\cal D})\to \Omega^1 ({\cal D})$ is the Lie derivative 
along it, we have 
\beq
L_Z\; \omega \wedge \omega =0.   
\label{symm}
\eeq
The interested reader is referred to Ref. \cite{bocharov} and in 
particular also to 
Ref. \cite{cerveau}, where 
the notions of homogeneous integrable Pfaffian forms and symmetry 
are found, and the corresponding integrating factor is deduced.\\
The definition of homogeneous 1-form we give below is the same as the 
one appearing in \cite{godbillonmec}. 
An homogeneous 1-form of degree $k+1$
\beq
\omega(x^1,\ldots,x^n)\equiv \sum_{i=1}^n\; \omega_i(x^1,\ldots,x^n)\; dx^i,
\eeq
is, by definition, such that the coefficients $\omega_i(x^1,\ldots,x^n)$ 
are homogeneous functions of degree $k$:
\beq
\omega_i(\lambda\; x^1,\ldots,\lambda\; x^n)=
\lambda^{k}\; \omega_i(x^1,\ldots,x^n) 
\ \forall\ i=1\ldots n.
\label{homog}
\eeq
In the case of an homogeneous 1-form 
the so-called radial vector field (which could be called also 
``Liouville field'' \cite{godbillonmec})  
\beq
Y\equiv \sum_{i=1}^n\; x^i\; \frac{\pa}{\pa x^i}
\label{liouv}
\eeq
is a symmetry for $\omega$. In fact, if $x$ stays for 
$x^1,\ldots,x^n$, one has that the flow of the 
vector field $Y$ is $\phi_t (x)= x\; \exp(t)$, thus 
\beq
L_Y\; \omega = (k+1)\; \omega.
\eeq
If the form is integrable and $\omega(Y)\not \equiv 0$, then 
$\omega(Y)$ is an integrating factor for $\omega$. 
This follows from the integrability condition 
\beq
\omega \wedge d \omega = 0
\label{integr}
\eeq
and from the definition of ``symmetry''. 
The condition  
$\omega(X)\not \equiv 0$ for a generic symmetry generator $X$ is 
equivalent to the requirement that the 
symmetry is nontrivial 
according to the definition of Ref. \cite{bocharov}, i.e., the 
vector field $X$ does not belong to the distribution $P$ of 
codimension one associated with the kernel of $\omega$. This means 
that the symmetry ``shuffles'' the leaves of the foliation 
associated with the integrable Pfaffian form $\omega$ but it leaves 
the foliation itself invariant (instead a trivial symmetry 
preserves each leaf) \cite{bocharov}. 
From a geometrical point of view, a trivial 
symmetry is tangent to the foliation, a nontrivial one is 
transverse to the foliation. 
Note that in Ref. \cite{cerveau}
a homogeneous Pfaffian form such that $\omega(Y)\not \equiv 0$ 
is called {\sl non-dicritical}.  
We sketch here the proof that $\omega(Y)$ is an integrating factor 
for the sake of completeness; 
an equivalent but more elementary proof, based 
on a ``by hand'' verification, is found in Appendix A. 
$f$ is an integrating factor for $\omega$ if 
\beqn
d\left(\frac{\omega}{f}\right)=0,
\eeqn
that is 
\beqn
f\; d\omega- df\wedge \omega=0.
\eeqn
We verify that $\omega(Y)$ satisfies the latter equation:
\beqna
&&\omega (Y) d\omega - (d\omega (Y))\wedge \omega = \\
&&(i_Y(\omega)) d\omega - ((d i_Y)(\omega))\wedge \omega = \\
&&(i_Y(\omega)) d\omega - (L_Y \omega)\wedge \omega +(i_Y(d\omega)) 
\wedge \omega = \\
&&i_Y(\omega\wedge d\omega)=0
\eeqna
where we have introduced the standard contraction operator 
$i_Y:\Omega^1 ({\cal D})\to \Omega^0 ({\cal D})$ and we have 
used (\ref{symm}),(\ref{integr}) 
and also standard identities 
($d\omega (Y)=(d i_Y)(\omega)=(L_Y-i_Y d)(\omega)=
L_Y\; \omega - i_Y(d\omega)$ and $i_Y (\omega \wedge d\omega)=
(i_Y(\omega)) d\omega+((i_Y d)(\omega))\wedge \omega$).

\section{thermodynamic entropy revisited}

In the framework of thermodynamics of homogeneous systems, 
the domain ${\cal D}$ of $\delta Q_{rev}$ is assumed to be an 
open convex set 
(in the discussion of 
the third law the boundary $T=0$ is introduced). Convexity of ${\cal D}$ 
is related to the request of a concave entropy 
(see the following subsections). 
From a mathematical point of view, homogeneity forces the domain ${\cal D}$ 
to be closed under multiplication by a positive real scalar 
$\lambda$, i.e. for each $(U,V,X^1,\ldots,X^n)\in {\cal D}$ it 
has to hold 
$(\lambda U,\lambda V,\lambda X^1,\ldots,\lambda X^n)\in {\cal D}$. 
This means that ${\cal D}$ has to be a cone, thus it has to be a 
convex cone. Given a convex set ${\cal C}$, a convex cone can 
be easily constructed, in fact the set\\  
${\cal K}_c=\{(\lambda U,\lambda V,\lambda X^1,\ldots,\lambda X^n)|
(U,V,X^1,\ldots,X^n) \in {\cal C}, \lambda>0\}$\\ 
is the smallest convex cone 
containing ${\cal C}$ and it has the property to be closed under 
addition (see \cite{rockafellar}, pp. 13-14). Notice that ${\cal D}$ 
is also a $C^\infty$ differentiable manifold.\\ 
The Pfaffian form $\deq$ is assumed to be of class at least 
$C^1({\cal D})$, which is a rather general condition 
for the validity of Frobenius' theorem. [According to Ref. 
\cite{liebyng}, a more general setting should be allowed, where 
the intensive variables are only locally Lipschitz-continuous. 
This requirement, which is physically well-grounded \cite{liebyng}, 
would require a more general setting for Frobenius' theorem. In 
fact, the Pfaffian form $\deq$ would be locally Lipschitz-continuous, 
which implies that $d(\deq)$ exists only almost everywhere, 
thus also the integrability condition is defined only a.e. 
We don't deal with this problem herein]. 
Then the  intensive variables 
$p(U,V,X^1,\ldots,X^n)$ and $\xi_i(U,V,X^1,\ldots,X^n)$, 
for $i=1,\ldots,n$, belong to $C^1({\cal D})$ (at least); they are 
supposed to 
be known functions of the extensive variables.\\
The Liouville operator is 
\beq
Y=U\; \frac{\pa}{\pa U}+V\; \frac{\pa}{\pa V}+
\sum_i\; X^i\; \frac{\pa}{\pa X^i}.
\eeq
\\ 
{\bf h2}) {\sf We require that the homogeneity symmetry of $\deq$ is 
nontrivial}.\\ 
\\
Then, 
an integrating factor for $\deq$ can be 
immediately deduced from the homogeneity property, and 
it is given by 
\beq
f(U,V,X^1,\ldots,X^n)\equiv i_Y(\omega)=U+p\; V-\sum_{i}\; \xi_i\; X^i \not \equiv 0.
\label{inteffe}
\eeq 
As a consequence, 
\beq
\frac{\delta Q_{rev}}{f}
\label{exdif}
\eeq
is a closed and, moreover, exact 1-form (being  
the domain simply connected). 
Thus 
\beq
\frac{dU+p\; dV-\sum_{i}\; \xi_i\; dX^i}{U+p\; V-\sum_{i}\; \xi_i\; X^i}
\equiv d\hat{S}
\label{hatsen}
\eeq
where $\hat{S}$ is a potential for the form. 
The 1-form $d\hat{S}$ is an homogeneous function of degree zero, 
as it follows from (\ref{hatsen}); in fact, the integrating factor $f$ is an 
homogeneous function of degree one:
\beqn
Y\; f = Y\; i_Y (\omega)=i_Y L_Y \omega = i_Y \omega =f
\eeqn
(the Cartan formula $L_X i_Y-L_Y i_X=i_{[X,Y]}$ holds), 
and the infinitesimal heat 
exchanged reversibly is an homogeneous function of degree one. The exact 
Pfaffian form in (\ref{hatsen}) has coefficients $(1/f,p/f,-\xi_1/f,
\ldots,-\xi_n/f)$ which 
are homogeneous functions of degree $-1$. Then $\has$ can be found only 
by quadratures (cf. \cite{ince}, p. 16-20).  
One has
\beq
\hat{S}(U,V,X^1,\ldots,X^n)-\hat{S}(U_{0},V_{0},X^1_{0},\ldots,X^n_{0})= 
\int_{\Gamma}\; 
\frac{dU+p\; dV-\sum_{i}\; \xi_i\; dX^i}{U+p\; V-\sum_{i}\; \xi_i\; X^i}
\label{hesse}
\eeq
where $(U_{0},V_{0},X^1_{0},\ldots,X^n_{0})$ is a reference state and 
$\Gamma$ is any reversible path connecting the reference state to the 
state $(U,V,X^1,\ldots,X^n)$ of interest [a path is a  
oriented curve in the thermodynamic domain 
which is at least piecewise $C^1$; one could also require, 
without loss of generality that the path is also simple, 
that is, non self-intersecting. 
A path, being defined along equilibrium states of the thermodynamic 
manifold, is, as a consequence, reversible in the sense of Refs. 
\cite{landsberg,callen}].\\
\\ 
{\bf h3}) {\sf We require that 
the thermodynamic foliation is defined by the leaves $\has=$ const.  
everywhere in ${\cal D}$.}\\ 
\\
As a consequence, 
$\has$ is required to be a well-defined 
function for each state belonging to the thermodynamic domain 
${\cal D}$. This assumption is natural in our framework; its 
necessity emerges in light 
of the discussion of the following subsections, in particular 
of subsections \ref{corgibbs} and \ref{poscon}.\\ 
\\ 
{\bf h4}) {\sf We require that the integrating factor $f$ is 
non-negative}\\ 
\\
The non-triviality condition $f\not \equiv 0$ is enforced by requiring 
that $f$ is of definite sign [one could also ask for $f\leq 0$, the 
choice $f\geq 0$ is made on a conventional basis]. 
The relation of hypothesis h4) with thermodynamics is evident 
from the following subsection, being related to the non-negative 
definiteness  of the absolute temperature.

\subsection{extensive entropy}

The potential $\has$ is related to the logarithm of a positive definite 
extensive potential $H$  
\beq
\has-\has_0=\log\left(\frac{H}{H_0}\right),
\eeq
where $Y\; H=H$ and $H_0\equiv H(U_{0},V_{0},X^1_{0},\ldots,X^n_{0})$ 
is the value of $H$ at the reference state 
$(U_{0},V_{0},X^1_{0},\ldots,X^n_{0})$. Analogously, 
$\has_0\equiv \has (U_{0},V_{0},X^1_{0},\ldots,X^n_{0})$.  
\\
In fact, one has 
\beq
d\; L_Y\; \hat{S} = L_Y\; d\hat{S} = L_Y\; \frac{\omega}{f}=0,
\eeq
being $L_Y\; (\omega/f) = (L_Y\; \omega)/f-(\omega\;/f^2)\;  (Y\; f)= 0$. 
Then 
\beq
L_Y\; \hat{S} = Y\; \hat{S} = q
\label{eqS}
\eeq
where $q$ is a constant. We can, without loss of generality, 
define another positive definite 
function $H$ such that 
\beq
\hat{S}\equiv \log(H).
\eeq
Then we get that (\ref{eqS}) is equivalent to the 
following equation for $H$:
\beq
Y\; H=q\; H,
\eeq
which means that 
$H$ is an homogeneous function of degree $q$.  
One has
\beq
d\hat{S}=\frac{dH}{H}=\frac{\omega}{f};
\eeq
as a consequence, 
\beq
dH=\frac{H}{f}\; \omega
\label{omeh}
\eeq
which is an homogeneous of degree $q$ closed Pfaffian form. 
Let us assume that $q\not =0$, i.e., that $H$ is not intensive.  
Then, from (\ref{pothom}) in Appendix A it follows
\beq
H = \frac{1}{q}\; \frac{H}{f}\; i_Y(\omega) = \frac{1}{q}\; H,
\eeq
which implies $q=1$. The case $q=0$ is treated in Appendix B, and 
it is excluded, $H$ cannot be intensive. See also \cite{qobh}, where 
a more general proof is given.\\ 
The leaves $\has=$ const. of the thermodynamic foliation 
coincide then with the leaves $H=$ const. and $H$ 
appears to play a privileged role in the construction of 
the thermodynamic foliation. 

\subsection{metrical entropy $S$}

We know that there exists a 
function $H$ such that 
\beq
\hat{S}(U,V,X^1,\ldots,X^n)-\hat{S}(U_{0},V_{0},X^1_{0},\ldots,X^n_{0})
=\log \left( \frac{H(U,V,X^1,\ldots,X^n)}{H(U_{0},V_{0},X^1_{0},\ldots,X^n_{0})} \right) 
\eeq
where $H$ is an extensive function. We have
\beq
\delta Q_{rev}=f\; d\has\; = \frac{f}{H}\; dH,
\eeq 
where $f/H$ appears as a new integrating factor and $H$ as 
extensive potential. By symmetry, the function $H$ is 
suggested to be the thermodynamic potential one is 
looking for, i.e. the so-called metrical entropy \cite{buchdahl}, 
because one can write
\beq
\frac{f}{H}\; dH=dU+p\; dV-\sum_{i}\; \xi_i\; dX^i
\eeq
where $f/H$ is intensive as each coefficient of the 
Pfaffian form is, and $H$ is extensive as each independent 
thermodynamic variable $U,V,X^1,\ldots,X^n$ is. 
This identification 
of $H$ with the metrical entropy is correct.   
We know, from the standard approach 
\cite{buchdahl,beattie,bazarov,wilson,landsberg}, 
that 
a metrical entropy has the property to be {\sl additive}. 
In particular, the entropy $S$ of a system composed by two subsystems 
in thermal equilibrium is $S=S_1+S_2$. Moreover, it is expected 
that the entropy of a system composed by $n$ replicas of a subsystem 
$\Sigma_0$ is $S=n\; S_0$, that is, it is {\sl extensive}. We just have 
the extensivity property for $H$. Moreover, also on a statistical 
mechanical ground, $S$ is required to be non-negative. 
Then, we can introduce on this ground the following assumption:\\
\\ 
{\bf h5}) {\sf We require that the metrical entropy $S$ is 
extensive}.\\ 
\\
We can easily show that $H$ is a metrical 
entropy because, up to a multiplicative constant, 
there exists only one extensive potential $S$ and only one 
intensive integrating factor $T$ such that $\deq=T\; dS$; 
as a consequence, $f/H=(\pa H/\pa U)^{-1}$ 
plays the role of absolute temperature. The proof is 
the following. We have obtained 
\beq
\omega=g\; dH,
\eeq
where $g\equiv f/H$ is homogeneous of degree zero. One may wonder if 
$g$ and $H$ are unique. By introducing another function $G(H)$  
we can obtain
\beq
\omega=\frac{g}{dG/dH}\; dG.
\eeq
We want again the new integrating factor 
$\bar{g}\equiv g/(dG/dH)$ to 
be homogeneous of degree zero and $G$ to be homogeneous 
of degree one. Then we have 
\beq
Y\; G=G
\eeq
and 
\beq
Y\; G(H)=\frac{dG}{dH}\; Y\; H=\frac{dG}{dH}\; H.
\eeq
Then 
\beq
G=\frac{dG}{dH}\; H
\eeq
which means
\beq
G= \gamma\; H
\eeq
where $\gamma$ is a constant. As a consequence, one finds 
\beq
\frac{1}{\bar{g}}=\frac{\pa G}{\pa U}=\gamma\; \frac{\pa H}{\pa U}= 
\gamma\; \frac{1}{g}.
\eeq
The entropy $H$ and the absolute temperature $g$ 
are then unique apart from a scale factor $\gamma$ which can be 
fixed by fixing the absolute temperature scale \cite{buchdahl}. 
$\has$ plays instead the role of 
empirical entropy \cite{buchdahl}.\\
We have found that, apart from a scale factor, 
\beqnl
H &\equiv& S,\\
\frac{\pa H}{\pa U} &\equiv& \frac{1}{T}=\frac{H}{f}
\eeqnl
The latter equality, together with the standard choice for $T$ to 
be non-negative, justifies h4).

\subsection{corroboration from Gibbs' approach}
\label{corgibbs}

$\has$ can also be easily related to the standard definition 
of entropy also by simply appealing to Gibbs' approach. 
In fact, it is sufficient to consider 
the fundamental equation of thermodynamics in 
the entropy representation for an homogeneous system \cite{callen}, which 
implies 
\beq
T\; S=U+p\; V-\sum_{i}\; \xi_i\; X^i;
\eeq
then
\beq
f=T\; S
\label{effe}
\eeq
this shows that $f$ cannot identically vanish in thermodynamics, 
thus, a posteriori, 
the integrable Pfaffian form $\deq$ has to be non-dicritical, 
and the homogeneity symmetry has to be non-trivial (or transversal). 
Moreover, 
\beq
d\hat{S}\equiv \frac{dS}{S}=d \log(S).
\label{ens}
\eeq
Of course, both (\ref{effe}) and (\ref{ens}) could be easily found 
at first sight from (\ref{inteffe}), but 
we have shown that on the ground of simple assumptions 
one can recover $S$ and $T$ as derived quantities without 
referring a priori to Gibbs' fundamental equation in the 
entropy representation.\\ 
We find that
\beq
\int_{S_0}^S\; \frac{dS}{S}= \int_{\Gamma}\;\frac{\omega}{f}.
\label{intess}
\eeq
As a consequence, 
\beq
\has-\has_0= \int_{\Gamma}\;\frac{\omega}{f}=\log(\frac{S}{S_0})
\label{hass}
\eeq
and 
\beqnl
S&=&S_0\; \exp (\hat{S}-\hat{S}_0)\\
&=&S_0\; \exp (\int_{\Gamma}\;\frac{\omega}{f}).
\label{posen}
\eeqnl
Notice that, a priori, one should write $d\has=d\log(|S|)$, and insert 
an absolute value in the argument of the logarithm in (\ref{hass}). 
In fact, from the point of view of classical thermodynamics, there is 
no constraint for $S$ to be non-negative.  
Nevertheless, the left side of (\ref{intess}) has to be of definite sign, 
contrarily $\has$ would not be well-defined where $S=0$ (the left side 
of (\ref{intess}) could still be defined as a principal value integral) 
and the thermodynamic foliation defined by $\has$ would not be 
in a one-to-one correspondence with the one defined by $S$, because 
the map $S \mapsto \has$ would not be injective.\\
\\
From (\ref{hass}) we can also deduce that $\has$ is not an homogeneous 
function, in fact a differentiable homogeneous function 
$g(U,V,X^1,\ldots,X^n)$ of degree $\alpha$ should satisfy $Y\; g=\alpha\; g$. 
For $\has$ we get 
\beq
L_Y\; \has = Y\; \has = 1.
\eeq

\subsection{positivity and concavity of entropy}
\label{poscon}

It is interesting to notice that  
{\sl the thermodynamic entropy $S$ is positive}. There is no 
need of restricting the range of the metrical entropy $S$ to 
positive values, because  the thermodynamic 
potential $H$, which is realized to be a metrical entropy, is, by 
definition, positive definite. 
Of course, this agrees with the statistical mechanical approach. 
The classical ideal gas violates this property, in fact  
its entropy becomes negative and goes to $-\infty$ as $T\to 0$; but  
this behavior being corrected at low temperatures by quantum mechanics. 
[The positivity of $S$ can be obtained in the framework we sketched in the 
previous subsection by recalling that, under the assumption h3), 
$\has$ is required to be well-defined on each thermodynamic 
state belonging to the thermodynamic domain where the 
thermodynamic foliation is constructed; see the discussion in 
the previous subsection]. 
States such that $S=0$ 
are singular and can be allowed to belong at most to the boundary 
of the thermodynamic domain, as we show in the following subsection. 
\\   
\\
As far as the concavity of the entropy is concerned, we limit 
ourselves to discuss how this fundamental stability requirement 
which is introduced in Gibbs' approach constraints the 
Pfaffian form $\deq$. Concavity of $S$ ensures that the system 
is thermodynamically stable \cite{callen}. We introduce 
then the assumption\\
\\ 
{\bf h6}) {\sf We require that the metrical entropy $S$ is 
concave}.\\ 
\\
Notice that the requirement of superadditivity for $S$ would be 
equivalent, in fact superadditivity and homogeneity of $S$ imply 
concavity \cite{galg,land}. Cf. also \cite{qobh}.\\ 
First, we underline that 
the concavity of $S$ is {\sl not} equivalent to the concavity 
of $\hat{S}$. 
From one hand, 
the concavity of $S$ is a sufficient condition for the concavity of 
$\has$; in fact, the logarithm of a concave function is a concave 
function (if $g$ is a concave 
function and $\phi$ is a concave non-decreasing function, then 
$\phi\circ g$ is a concave function. The logarithm is a concave 
non-decreasing function). The concavity of $S$ is not a necessary 
condition for the concavity of $\has$ 
(if one considers $S=\exp(-x^2)$, which is not concave on $\RR$ 
but only for $2\; x^2 -1<0$, 
one still obtains $\has=-x^2$ which is concave on $\RR$). 
On the other hand, the concavity of $\has$ does not ensure the 
concavity of $S$, because, in general is not true that the 
exponential of a concave function is concave (e.g., let us 
consider again $\has=-x^2$, which is a concave function on $\RR$. 
The function $S\equiv \exp(-x^2)$ 
is not concave on the whole real line). 
\\
The concavity of $S$ can be obtained by imposing suitable conditions 
on the coefficients of the Pfaffian form $\deq$. These conditions can 
be deduced from the standard ones on $S$ \cite{callen}. Let us 
consider for simplicity $S(U,V,N)$. We get 
\beqnl
\frac{\pa S}{\pa U} &=& S\; \frac{1}{f}\\
\frac{\pa S}{\pa V} &=& S\; \frac{p}{f}\\
\frac{\pa S}{\pa U} &=& S\; \frac{-\mu}{f}.
\eeqnl
The Hessian matrix for $S$ is easily deduced to be 
\[ D^2 S\equiv \frac{S}{f^2}\; \left[
\begin{array}{ccccc}
1-\frac{\pa f}{\pa U} & 
p-\frac{\pa f}{\pa V} & 
-\mu-\frac{\pa f}{\pa N}
\cr
&&\cr
p-\frac{\pa f}{\pa V} & 
p^2+f\; \frac{\pa p}{\pa V}- p\; \frac{\pa f}{\pa V} &
-\mu\; p+ f\; \frac{\pa p}{\pa N}-p\; \frac{\pa f}{\pa N}   
\cr
&&\cr
-\mu-\frac{\pa f}{\pa N} &
-\mu\; p+ f\; \frac{\pa p}{\pa N}-p\; \frac{\pa f}{\pa N} &
\mu^2 - f\; \frac{\pa \mu}{\pa N}+\mu\; \frac{\pa f}{\pa N} 
\end{array} 
\right].
\]
Concavity requires that all principal minors of odd order should 
be negative and all principal minors of even order should be 
positive. The Hessian determinant is zero because of the homogeneity 
of $S$. Then, in order to obtain a concave $S$ one has simply to 
impose on the homogeneous integrable Pfaffian $\deq$ 
the following (necessary and sufficient) conditions involving 
the coefficients of the Pfaffian form and the integrating factor $f$: 
\beqnl
&&1-\frac{\pa f}{\pa U}<0\\
&&(1-\frac{\pa f}{\pa U})
(p^2+f\; \frac{\pa p}{\pa V}- p\; \frac{\pa f}{\pa V})-
(p-\frac{\pa f}{\pa V})^2>0.
\eeqnl
Notice that $1-\pa f/\pa U<0$ amounts simply to 
$\pa T/\pa U>0$, which is simply the positivity  of the heat capacity 
at constant $V,N$.

\subsection{conditions for $\has$ and $f$ to be globally defined}

The homogeneity 
of the integrable non-dicritical Pfaffian form $\deq$ 
allows to find an integrating factor $f$ whose expression holds globally 
by construction (by hypothesis,  
$p(U,V,X^1,\ldots,X^n)\in C^1({\cal D})$ and 
$\xi_i(U,V,X^1,\ldots,X^n)\in C^1({\cal D})$ for all $i=1,\ldots,n$).  
A priori it is not possible to ensure that also $\has$ is defined 
everywhere, because singularities can arise where $f=0$, i.e., if the set 
of zeroes of the integrating factor 
$Z(f)=\{(U,V,X^1,\ldots,X^n)|\; f(U,V,X^1,\ldots,X^n)=0\}$ is non-empty. 
In fact, where $f$ vanishes $\has$ can be not well-defined, as it is 
clarified in the following, in contrast with 
our requirement that the thermodynamic foliation is described by the 
potential $\has$ in the convex set ${\cal D}$. 

\subsubsection{the set $Z(f)$}

We know that 
$f=T S\geq 0$, 
and obviously   $Z(f)=Z(T)\cup Z(S)$, where $Z(T),Z(S)$ are the 
sets where $T,S$ vanish respectively. 
Surely $\has$ is not well-defined for 
each state belonging to $Z(S)$, because of (\ref{hass}). 
Thus, one has to require that 
$Z(S)$ belongs at most to the boundary of ${\cal D}$. This requirement 
is corroborated by the analysis of convex functions \cite{belg30}, in fact 
$I\equiv -S$ is a convex function which attains its maximum 
value in $Z(S)$ if $Z(S)\not = \emptyset$ 
(recall that $S\geq 0$ in our framework), and, 
in order that $I$ is non-constant, it is necessary that $Z(S)$ 
is contained in the boundary of the convex domain ${\cal D}$ 
[cf. also \cite{roberts}, thm. C p. 124].  
$Z(T)$ is contained in the boundary as well (as it is evident 
if $T$ appears as independent variable). Then also $Z(f)$ is 
contained in the boundary of the thermodynamic domain.\\
Moreover,\\ 
\\
{\bf h7}) 
{\sf We require that $Z(S)\subseteq Z(T)$.}\\ 
\\ 
In the following discussion, we indicate with $z$ collectively the 
independent thermodynamic variables. The assumption h7) 
is introduced because 
any state $z$ such that $T_z>0$ and $S(z)=0$ should belong to the 
boundary of the thermodynamic domain and 
should have the 
peculiar property to allow the system only to absorb heat along 
any non-adiabatic path $\gamma_z$ starting from $z$ in a neighborhood 
$W_{z}$ of $z$. In fact, 
let us define the heat capacity along a 
path $\gamma:[T_0,T]\to {\cal D}$ which does not include 
isothermal sub-paths:
\beq
C_{\gamma}(T)\equiv T\; \left(\frac{dS}{dT} \right)_{\gamma}=
\omega(\dot{\gamma}). 
\eeq
If $\gamma_z$ is a path starting from the state $z$, then   
\beq
S(y)=\int_{T_z}^{T_y}\; \frac{dT}{T}\; C_{\gamma_z}(T)
\eeq
should be positive for any state $y$ non isoentropic to 
$z$ in $W_{z}$, because $S(y)>S(z)=0$,  
which is possible only for heat absorption (in fact, 
$C_{\gamma_z}(T)<0$ would be allowed for states such that 
$T_z<T_y$, which would imply heat absorption, and  
$C_{\gamma_z}(T)>0$ would be allowed for states such that 
$T_z>T_y$). [Heat absorption should occur also for any 
isothermal path starting from the state $z$, in fact 
$Q=T_z\; \Delta S$ along an isothermal path and $\Delta S>0$]. 
Thermal contact with a colder body at 
$T<T_z$ should allow an heat flow outgoing from the system 
(see also the discussion in Ref. \cite{belg30}). 
Then, no quasi-static approximation 
of such a thermal contact can be allowed, no matter how near 
to $T_z$ the temperature of the colder body could be, 
because, at least in a neighborhood of $T_z$, the system 
could only absorb heat. This behavior can be considered 
pathological, and the occurrence of the absolute minimum $S=0$ of 
the thermodynamic entropy at $T>0$ is refused in the 
framework of standard thermodynamics. 
As a consequence of the rejection of states with $S=0$ at $T>0$ 
one finds $Z(f)=Z(T)$. Notice that in the 
classical ideal gas case, where $S$ is allowed to become negative, 
one finds that $Z(f)\supset Z(T)$ 
because $T$ vanishes for $U=0$ and $S$ vanishes before 
the hypersurface $U=0$ is reached.\\

Notice that h7) is 
automatically implemented if the boundary $T=0$ is described by 
a (maybe even smooth) 
function $U=b(V,X^1,\ldots,X^n)$ and conditions ensuring the 
continuity of $S$ at $T=0$ are allowed \cite{belg30}. 
The function $b(V,X^1,\ldots,X^n)$ can be construed as a 
ground-state energy. h7) can be substituted 
by the stronger assumption\\ 
\newpage
\noindent
$\overline{{\mathrm {\bf h7}}}$) 
{\sf We require that $T=0$ is described by  $U=b(V,X^1,\ldots,X^n)$, 
where $b$ is an extensive convex function defined on a convex cone 
${\cal K}_b$, and that the domain ${\cal D}\cup \pa {\cal D}$ 
coincides with the 
epigraph of $b$:
\beq
{\cal D}\cup \pa {\cal D}\equiv {\mathrm epi}(b)=
\{ (U,V,X^1,\ldots,X^n)\; |\; (V,X^1,\ldots,X^{n})\in 
{\cal K}_b, U\geq b(V,X^1,\ldots,X^n) \}.
\eeq}
\\
As a consequence, by defining the extensive coordinate 
$B\equiv U-b(V,X^1,\ldots,X^n)\geq 0$ it is easy to show that 
$Z(S)\subseteq Z(T)$, because  for an 
everywhere continuous entropy it holds 
\beq
S(B,V,X^1,\ldots,X^{n})=S(0,V,X^1,\ldots,X^{n})+
\int_0^B\; dY\; \frac{1}{T(Y,V,X^1,\ldots,X^{n})}, 
\eeq
where $S(0,V,X^1,\ldots,X^{n})$ is the value attained by $S$ at $B=0$ 
by continuity [all the matematical properties ensuring the existence of the 
improper integral $\int_0^B\; dY\; 1/T$ are implicitly assumed]. 
See \cite{belg30} for further details. Then we have $Z(f)=Z(T)$. 
Notice that, if 
$b\equiv 0$, then 
${\cal D}\cup \pa {\cal D}={\cal K}\times \overline{\RR_+}$, where 
${\cal K}\ni (V,X^1,\ldots,X^n)$ is a convex cone.\\  
\\ 
Then $\has$ and $S$ are defined everywhere, because  $\deq/f$ 
is a closed Pfaffian form defined everywhere on ${\cal D}$. 
This solves 
the problem of ensuring the global existence of the integrating 
factor and of the potential, in particular, the entropy and the 
temperature are defined globally (about this problem in the 
frame of Carath\'eodory approach, cf. \cite{bernstein,boyling}). 

We further introduce the following assumption:
\\ 
\\
{\bf h8}) 
{\sf We require that to each level set $S=$ const. corresponds a unique 
leaf.}\\
\\ 
In general, $S$ is a submersion whose leaves are the connected 
components of $S^{-1}(c)$, where $c\in \RR_+$ is a constant. 
From a physical point of view, it is expected that each isoentropic 
surface $S=c=$ const. is path-connected, in fact, given a state 
$X$ and a state $Y$ lying on the same isoentropic surface of an homogeneous 
thermodynamic system, it is physically mandatory that there exists an 
adiabatic reversible transformation [i.e. an adiabatic path] 
which connects $X$ and $Y$. Thus, assumption h8) has to be ensured. 
In Appendix \ref{hyph8} we show a possible implementation of 
assumption h8).\\

It can be noted that the conditions $\omega(Y)\not \equiv 0$ and 
$\omega(Y)=0$ only on the boundary of the thermodynamic domain ensure 
that, given a fixed reference point $z_0\equiv (U_0,V_0,X^1_0,\ldots,X^n_0)$, 
the integral curve $\gamma_{z_0}(t)$ of the vector field $Y$  
meets each leaf $S=$ const. of the foliation if h8) is implemented, 
in fact $S({\cal D})=(0,\infty)$ and it holds 
\beq
(\gamma_{z_0}^{\ast}(t) S) (U_0,V_0,X^1_0,\ldots,X^n_0)=S(\exp(t)\; U_0,
\exp(t)\; V_0,\exp(t)\; X^1_0,\ldots,X^n_0)=\exp(t)\; 
S(U_0,V_0,X^1_0,\ldots,X^n_0).
\eeq
Moreover, there exists only one intersection with each leaf. 
The given curve is then transverse with respect to the foliation.\\

Another physically well-grounded assumption is\\
\\  
{\bf h9}) 
{\sf We require that $\frac{\pa S}{\pa V}$ is positive.}\\
\\ 
This requirement means simply that the pressure $p$ is positive definite.

\subsection{the problem of $T=0$}

This kind of approach is very interesting also from the point of 
view of the third law of thermodynamics 
\cite{buchdahl,beattie,bazarov,landsberg,kestins}. A more complete analysis 
of this topic is the subject of Ref. \cite{belg30,belg31}. We limit 
ourselves to a short summary of some points which have 
an evident link with the subject of this paper, referring the 
reader to Ref. \cite{belg30,belg31} for details, for 
a general discussion and for also for complete references to the 
literature on the third law.\\ 

We assume that the 
$T=0$ is a connected hypersurface (see also \cite{belg30}), 
which coincides with the 
adiabatic boundary $\deq=0$ of the thermodynamic domain, i.e.\\ 
\\  
{\bf h10}) 
{\sf We require that $T=0$ is a connected integral manifold of $\deq$.}\\
\\ 
This means that, if $\overline{\mathrm h7}$) is implemented,   
$U=b(V,X^1,\ldots,X^n)$ has to be a solution of the 
Mayer-Lie system which is equivalent to the Pfaffian equation $\omega=0$ 
(see \cite{frankel}), i.e. the following conditions 
\beqnl
\frac{\pa U}{\pa V}&=&-p(U,V,X^1,\ldots,X^n)=\frac{\pa b}{\pa V}\\
&&\cr
\frac{\pa U}{\pa X^i}&=&\xi_i (U,V,X^1,\ldots,X^n)=\frac{\pa b}{\pa X^i}
\quad \hbox{for}\ i=1,\ldots,n
\eeqnl
have to be implemented. Then the function $b$ has to be at least of 
class $C^1 ({\cal K}_b)$.\\ 
Requirement h10) is not necessary if $\overline{\mathrm h7}$) is 
implemented and $f$ and $\omega$ are $C^1$ everywhere (then $b$ has to be 
of class $C^2 ({\cal K}_b)$). In fact, from 
$f\; d\omega=df\wedge \omega$ one finds that the pull-back of the 
map $F_b:{\cal D}\to \pa {\cal D}$, where $F_b^{\ast}(f)=f\circ F_b=0$, 
implements $0=F_b^{\ast}(df\wedge \omega)=F_b^{\ast}(df)\wedge 
F_b^{\ast}(\omega)$. This implies that there exist a real function $h\not=0$ 
defined in $\pa {\cal D}$ 
such that $F_b^{\ast}(\omega)=h\; F_b^{\ast}(df)$ and 
$F_b^{\ast}(df)=0$ because $F_b^{\ast}(f)=0$. Cf. also \cite{cerveau}. 
Under more general 
conditions on $f$ and $\omega$ it can be non-trivial that $f=0$ 
is an integral manifold.

The use $T$ as explicit independent coordinate is not suitable for 
studying the nature of the integral submanifold $T=0$, 
in fact the Pfaffian form $\deq$ 
seems to be singular in $T=0$ (the point is that the map $U\mapsto T$ 
is not a diffeomorphism in $T=0$). In order to avoid this problem  
the explicit use of $U$ or $B$ [or of another regular coordinate] is 
mandatory. 
We then use $B$ explicitly. See also 
the discussion in \cite{belg30,belg31}. 
Outside $T=0$ [i.e. $B=0$] there exists a thermodynamic foliation whose leaves 
are the hypersurfaces $S=$ const. The nature of $T=0$ is ambiguous, 
in the sense that it can correspond to a leaf of the thermodynamic 
foliation, the leaf $T=0$, or it simply can correspond to an integral 
manifold of $\deq$ which is intersected by other integral manifolds 
of $\deq$.  
In both cases, $S:{\cal D}\to \RR_+$ 
represents an at least $C^2({\cal D})$ submersion which  generates the 
thermodynamic foliation at $T>0$. We can show that the validity of the 
third law of thermodynamics can be related with the geometric 
nature of the integral submanifold $T=0$. 

In fact, under the hypothesis  
that the entropy $S$ is concave and continuous at the surface $T=0$, in the 
sense that there exists the limit as $B\to 0^+$ of $S$ for any finite 
value of the parameters $V,X^1,\ldots,X^n$ 
[notice that the theory of convex functions allows to 
define $S$ at $T=0$, i.e. when $U=b$ \cite{belg30}],  
it can be shown that $T=0$ corresponds to a special leaf of the 
thermodynamic foliation generated by the Pfaffian form 
$\deq$ if, and only if, the entropic version of the third law
\beq
\lim_{T\to 0^{+}} S=0
\label{planck}
\eeq
holds.  
We notice that Planck's restatement of the 
third law (\ref{planck}) is trivially mandatory for an 
homogeneous system, in fact  $\lim_{B\to 0^{+}} S(B,V,X^1,\ldots,X^n)=S_0$, 
with $S_0>0$ a positive constant, implies also that 
$S_0=\lim_{B\to 0^{+}} 
S(\lambda B,\lambda V,\lambda X^1,\ldots,\lambda X^n)=
\lambda \lim_{B\to 0^{+}} S(B,V,X^1,\ldots,X^n)=\lambda S_0$, which 
is possible only for $S_0=0$. It can also be noticed that 
the positivity and the concavity of $S$ forces $S$ to be finite 
as $B\to 0$ \cite{belg30}. In fact, the behavior of the convex 
function $I=-S$ at the boundary has to be such that 
\beq
\liminf_{x\to x_0}\; I(x) > -\infty
\eeq
for any $x\equiv (U,V,X^1,\ldots,X^n)$ which converges to 
$x_0\equiv (U_0,V_0,X^1_0,\ldots,X^n_0)$ belonging to 
the boundary of the convex domain 
[cf. problem F p. 95 of Ref. \cite{roberts}]. Then a positive 
$S$ has to be finite as $T\to 0$.\\
If (\ref{planck}) holds, then $Z(S)=Z(T)$ and  
no leaf $S=$ const. can intersect $T=0$, in fact only the isoentropic 
surface $S=0$ could intersect $T=0$.  
If (\ref{planck}) is violated, and 
$\lim_{B\to 0^{+}} S(B,V,X^1,\ldots,X^n)=S(0,V,X^1,\ldots,X^n)$, 
then $Z(S)\subset Z(T)$ and $T=0$ 
is not a leaf; in fact, one can uniquely define the entropy to be 
$S(0,V,X^1,\ldots,X^n)$ at $(B=0,V,X^1,\ldots,X^n)$, and a peculiar 
intersection of adiabatic surfaces occurs \cite{belg30}. 
The main point is that, if (\ref{planck}) is violated, a state belonging 
to the set $Z(T)$ can also belong to an isoentropic surface 
intersecting $T=0$, and 
there are two adiabatic paths starting from that state, one contained 
in $T=0$ and the other contained in the aforementioned isoentropic surface. 
This happens as a consequence of the fact that the surface 
$S=S(0,V,X^1,\ldots,X^n)=$ const. $>0$, if $(0,V,X^1,\ldots,X^n)$ is not 
a local minimum of $S$, necessarily intersects the 
graph of $S$ along a codimension one submanifold which is connected to 
the submanifold $T=0$. [The proof is found in \cite{belg30}; see also 
\cite{belg31}]. This means that the integral manifold 
$S=S(0,V,X^1,\ldots,X^n)=$ const. $>0$ reaches $T=0$ and is actually 
tangent to it \cite{belg30}. 
Notice that, given such a point $(0,V,X^1,\ldots,X^n)$, 
then each point $(0,\lambda\; V,\lambda\; X^1,\ldots,\lambda\; X^n)$ 
belonging to the cone of $(0,V,X^1,\ldots,X^n)$ is such that 
$S(0,\lambda\; V,\lambda\; X^1,\ldots,\lambda\; X^n)=\lambda\; 
S(0,V,X^1,\ldots,X^n)$ and it is not a local minimum as well [the 
proof is trivial]. Then, 
each (would-be) leaf $S=$ const. intersects $T=0$. 
One does not obtains a well-defined thermodynamic foliation if $T=0$ 
is included, actually a foliation can be obtained in this case only 
if zero-measure set represented by the surface $T=0$ is excised from 
the thermodynamic domain. 
Contrarily, one obtains a foliation except for a zero-measure set represented 
by the surface $T=0$, i.e. the inner part of the 
thermodynamic domain is foliated into surfaces $S=$ const., and locally 
the adiabatic inaccessibility holds, but these (would-be) 
leaves are actually connected by the 
adiabatic surface $T=0$ and the adiabatic 
inaccessibility is violated, even if only along special curves 
reaching $T=0$ \cite{belg30}. 
Note that 
the failure of the Lipschitz property for ordinary differential 
equation for adiabatic curves starting at $T=0$ is related to 
the singular behavior of the thermodynamic foliation at $T=0$ 
when (\ref{planck}) is violated \cite{belg30}. Note also that, on the 
ground of assumption h8), when (\ref{planck}) is violated one 
has to impose that any point $(0,V,X^1,\ldots,X^n)$ which is a local minimum 
has to be also a global minimum (contrarily, h8) would be violated).\\ 

If, instead, $T=0$ is a leaf, 
then no leaf $S=$ const. is allowed to reach $T=0$. Leaves 
$S=$ const. can at most asymptotically approach $T=0$, but no one can be 
extended therein. (\ref{planck}) has to hold for a continuous 
$S$ because the limit of $S$ as $T\to 0$ cannot depend on the 
parameters $V,X^1,\ldots,X^n$ (otherwise $T=0$ is not a leaf) 
\cite{belg30,belg31}.\\ 

Note that also the approach discussed herein 
is able to make immediately evident the necessity of the 
discussion of the states which belong to $Z(f)$, i.e., as a 
consequence of h7), of the states which belong to $Z(T)$. The third 
law has to be discussed in the framework of thermodynamic formalism, 
and the validity of (\ref{planck}) emerges as a ``regularity'' 
condition on the boundary $T=0$ 
for the differential equation $\deq=0$ of thermodynamics, in 
particular it appears as a condition which has to be imposed if 
a foliation of the whole thermodynamic domain (including $T=0$) 
has to be obtained.\\   

In the framework of our approach,  (\ref{planck}) 
holds if and only if, whichever path $\gamma^0$ is chosen in 
approaching $T=0$ and having a point belonging to $T=0$ as its final 
point, the integral of the form $\deq/f$ diverges to $-\infty$. 
The condition is clearly 
necessary, because (\ref{planck}) implies the divergence of 
(\ref{hass}); its sufficiency is also evident, being equivalent 
to $\log(S)\to -\infty$. 
Notice that this condition on the integral of $\omega/f$ 
does not require to define the entropy at $T=0$. 

Moreover, it is easy to show that a sufficient condition for 
(N) to hold is $\omega\in C^1$ everywhere [i.e. also on the boundary 
$T=0$], which implies $f\in C^1$ everywhere. In fact, 
one has
\beq
\frac{\pa f}{\pa U}=1+S\; \frac{\pa T}{\pa U}
\label{pafu}
\eeq
which has to be finite at $T=0$; 
it holds $\pa T/\pa U=1/C_{X^1,\ldots,X^{n+1}}$, where 
$C_{X^1,\ldots,X^{n+1}}$ is the 
standard heat capacity, which has to vanish in the limit 
as $T\to 0$ because $S$ has to be finite in that limit. 
As a consequence, $\pa T/\pa U=1/C_{X^1,\ldots,X^{n+1}}\to \infty$ 
as $T\to 0^+$, thus  
(\ref{pafu}) is finite only if $S\to 0$ as $T\to 0^+$.\\ 
Further details and conditions are the subject of 
forthcoming papers \cite{belg30,belg31}.

\subsection{a note on the reference state}
\label{refer}

The reference state $(U_{0},V_{0},X^1_{0},\ldots,X^n_{0})$ has to satisfy 
a constraint which is trivial but its discussion can be 
physically interesting. From the point of view of thermodynamics, 
given a system described by three variables $(U,V,N)$, 
reference states like $(0,0,0)$ or $(0,V,0)$ are meaningless. 
In fact, they correspond to the absence of the system under study. 
Statistical mechanics provides a criterion for obtaining a meaningful 
thermodynamic description: there should be a statistically relevant 
number of elementary constituents (atoms, molecules, particles) 
of the system. In particular, 
the thermodynamic limit is well-known to be the tool allowing to 
find out thermodynamics from statistical mechanics. [See also \cite{qobh}]. 
$N=0$ correspond to the absence of constituents. 
Without necessarily referring to the above statistical mechanical 
suggestion, one may wonder when a thermodynamic description 
starts being available. The system should be such that 
thermodynamic fluctuations allow to define 
meaningful macroscopic variables; moreover, the system 
under study should be realized to be involved with 
appropriate thermodynamic properties. In other terms, 
a ``preparation'' procedure of the system, in such a way that 
a macroscopic description becomes available and a set of 
relevant thermodynamic variables can be identified, should be considered 
as a preliminary condition to the thermodynamic description. 

\subsection{fundamental equation and homogeneous Pfaffian form}
\label{funpfa}

It is known that the fundamental equation 
contains all the relevant thermodynamic 
information about the system by construction \cite{callen}. 
In order to determine the fundamental equation $S$ 
by integration of the 
Pfaffian form $\delta Q_{rev}$, 
it is necessary to consider 
the Pfaffian form $\delta Q_{rev}$ representing a 
sufficiently 
general thermodynamic transformation which can be 
implemented with the system under study. Before clarifying 
what we mean by ``sufficiently general'', let us discuss 
some examples.  
In the case of the classical ideal gas, one should 
consider the system as open and let the extensive variables 
$U,V,N$ to appear in $\deq$. As it is easy to show, 
by considering $\deq$ with $N=$ const. (closed system) leads 
to a wrong fundamental equation, to be compared with 
the correct one displayed in Ref. \cite{callen}.  
If a photon gas is studied, only two variables are necessary, 
$U,V$, whereas $N$ is not a good parameter [which is related 
to the fact that the number of particles is not a good 
observable for a massless particles gas]. 
These examples show that neglecting some variable 
one would find incorrect results. Moreover,  
the relevant extensive variables should be settled out 
by considering, for each system, its own macroscopic features.  
Let us suppose that a system can be consistently described, when 
e.g. insulated, by means of three extensive variables $U,V,N$ 
appearing in the fundamental equation $S(U,V,N)$. This does 
not exclude a priori the possibility to consider the 
system in a more general situation in which the 
system is under the influence e.g. of some external field which 
could give rise to new thermodynamic features [as a paramagnetic gas 
under the influence of an external magnetic field, 
one could introduce an external magnetic field $\vec{B}_e$ 
and the magnetization $\vec{M}$ as in Ref. \cite{callen}]. 
Then one can allow an extension of the thermodynamic space, 
by enlarging the set of the extensive variables, e.g. by passing 
from $U,V,N$ to $U,V,N,X$, where $X$ is the new extensive 
variable which can assume the value $X=0$ and whose conjugate 
intensive variable is $\xi$. A new fundamental equation 
$\bar{S}(U,V,N,X)$ is obtained, which, by continuity, reduces 
to $S(U,V,N)$ when $X=0$: $\bar{S}(U,V,N,X=0)=S(U,V,N)$. 
Of course, an enlarged set of 
integrability conditions should be verified [cf. eqn. 
(\ref{integre}) in the next section]. A consistency condition for 
such an extension to hold consists in requiring that 
$\xi(U,V,N,X=0)=0$. This can be easily realized if one 
refer to the Massieu potentials which are obtained by performing a 
Legendre transformation of the fundamental equation \cite{callen}. If 
$\bar{\phi}(1/T,p/T,\mu/T,\xi/T)$ is the Massieu potential associated 
with $\bar{S}$ in 
which only intensive variables appear, then, in order to obtain 
a consistent reduction of thermodynamic degrees of freedom by 
posing $X=0$, it is necessary to put $\xi=0$.\\ 
Then ``sufficiently general'' means that 
in the expression of the Pfaffian form $\deq$ one takes into 
account all the extensive 
variables which cannot be consistently set equal to zero. 
This kind of approach to the problem of the construction of 
the fundamental equation appears to be more realistic than 
the one requiring to refer to the Pfaffian form $\deq$ 
corresponding to the most general transformation the system 
could be allowed to perform [the most general transformation 
is hardly known]. An explicit construction of this kind occurs 
in black hole thermodynamics, and it is discussed in Ref.   
\cite{qobh}. Because of the presence of gravity, thermodynamics 
is no more homogeneous, but a quasi-homogeneous symmetry 
of the Pfaffian form $\deq$ can still be identified 
\cite{qobh}. We point out also that quasi-homogeneity is the 
natural symmetry for standard thermodynamics when one or more 
independent variables are intensive \cite{beqo}.

\section{Comparison with the standard approach}

In the standard approach to homogeneous systems, in the so-called 
entropy representation, it is known that the knowledge of the fundamental 
equation is equivalent to the knowledge of the so-called state equations 
\cite{callen}. In particular, 
we have the following general scheme for 
reconstructing the fundamental equation of thermodynamics from 
the state equations\cite{callen}. This method is equivalent to 
the one we propose, where we know the intensive variables 
$p, \xi_1,\ldots,\xi_n$ and we reconstruct $S$. 
Given a system described by $n+2$ extensive variables, it is 
sufficient to know $n+1$ state equations, i.e. $n+1$ intensive 
functions (pressure, chemical potential,...) in order to 
reconstruct the $(n+2)$-esime state equation within an additive constant, 
and, then, to reconstruct the 
fundamental equation. In fact, being $S$ an homogeneous 
function of degree one of the extensive variables, one has
\beq
S= \frac{1}{T}\; U+ \frac{p}{T}\; V-\sum_{i} \frac{\xi_i}{T}\; X^i;
\eeq
moreover, by differentiating the above formula for $S$, 
the Gibbs-Duhem equation follows, which allows to 
reconstruct the $(n+2)$-esime state equation from the other ones:
\beq
U\;d\left(\frac{1}{T}\right)+V\; d\left(\frac{p}{T}\right)-\sum_{i}  X^i\; 
d\left(\frac{\xi_i}{T}\right)=0.
\eeq
Here, we assume to know $p,\xi_1,\ldots,\xi_n$ and we 
can recover $T$ (within an integration constant) from the 
knowledge of $p,\xi_1,\ldots,\xi_n$. 
Then we can reconstruct $S$. Of course, 
in general such a reconstruction of the $(n+2)$-esime state equation  
is possible only under suitable integrability conditions to be 
satisfied by $p,\xi_1,\ldots,\xi_n$. These integrability 
conditions coincide with the integrability conditions of $\deq$. 
A Pfaffian form 
\beq
\omega=\sum_i\; \omega_i\; dx^i
\eeq
is integrable if it satisfies the integrability condition 
$\omega\wedge d\omega=0$, that is, if, for any $i,j,k$ 
\beq
l_{ijk}=\omega_i\; \left(\frac{\pa \omega_j}{\pa x^k}-
\frac{\pa \omega_k}{\pa x^j}\right)+
\omega_j\; 
\left(\frac{\pa \omega_k}{\pa x^i}-\frac{\pa \omega_i}{\pa x^k}\right)+
\omega_k\; 
\left(\frac{\pa \omega_i}{\pa x^j}-\frac{\pa \omega_j}{\pa x^i}\right)=0. 
\label{integre}
\eeq
[See e.g. Ref. \cite{buchdahl}]. 
These conditions have to be satisfied by (\ref{heat}). E.g., in the 
case of three variables $U,V,N$, one has
\beq
l_{UVN}=\frac{\pa p}{\pa N}+\frac{\pa \mu}{\pa V}+\mu\; 
\frac{\pa p}{\pa U}-p\; \frac{\pa \mu}{\pa U}=0.
\label{intre}
\eeq
These conditions ensure that $\deq/f$ is an exact differential. 
Let us consider the Gibbs-Duhem equation in which 
$T$ is unknown 
\beq
d\; \log \left(\frac{1}{T}\right)= 
- \frac{V\; dp-\sum_i\; X^i\; d\xi_i}{U+p\; V-\sum_i\; X^i\; \xi_i}=
- \frac{V\; dp-\sum_i\; X^i\; d\xi_i}{f};
\label{gdt}
\eeq
the integrability conditions ensure that (\ref{gdt}) 
has a right member which is an exact differential, as can be easily 
verified. In fact, let us consider 
\beq
\frac{\deq}{f}=\frac{dU+p\; dV-\sum_i\; \xi_i\; dX^i}{f}=
\frac{df}{f}-\frac{V\; dp-\sum_i\; X^i\; d\xi_i}{f};
\eeq
it is evident that $d\has=\deq/f$ holds if and only if 
\beq
\frac{V\; dp-\sum_i\; X^i\; d\xi_i}{f}
\eeq
is the differential of a function. Moreover, from $f=T\; S$ and 
from
\beq
d\has=\frac{df}{f}-\frac{V\; dp-\sum_i\; X^i\; d\xi_i}{f}=\frac{dS}{S}
\eeq
the Gibbs-Duhem equation (\ref{gdt}) follows. 
Note that the Gibbs-Duhem equation (\ref{gdt}) allows to find 
$\log(1/T)$ within an additive constant, which means that 
$T$ is determined within a multiplicative constant, and 
also $S$ is determined within a multiplicative constant. In 
in this respect, the integration of $\omega/f$ and the 
integration of (\ref{gdt}) are equivalent.\\ 
As a counter-example where the integrability conditions fail, 
let us consider $\deq=du+p dV-\mu\; dN$, 
with $p=U/V$ and $\mu=U V/N^2$. The integrability condition (\ref{intre}) 
is not satisfied, in fact $l_{UVN}=U/N^2\not = 0$. One has 
\beqna
d\; \log \left(\frac{1}{T}\right)&=& 
\frac{V-N}{U (2 N - V)}\; dU+\frac{V+N}{2 N - V}\; dV-
\frac{2 V}{N (2 N - V)}\; dU\\ 
&\equiv& a_U\; dU+a_V\; dV+a_N\; dN.
\eeqna
The right member is not an exact differential, because e.g. 
$\pa_V a_U = N/U (2 N-V)^2\not = \pa_U a_V=0$. About the Gibbs-Duhem 
equations see also \cite{qobh} and \cite{beqo}.

\subsection{densities}

We introduce the so called 
{\sl densities}. One defines density the ratio of the extensive 
independent variables with respect to one of them. For example, 
it is possible to define:
\beqna
&&u\equiv U/V\\
&&x^i\equiv X^i/V.
\eeqna
The $n+1$ densities so obtained are the physical degrees of 
freedom in the case of an homogeneous substance. Then the 
fundamental equation in the entropy representation becomes
\beq
S(U,V,X^1,\ldots,X^n)=V\; s(u,x^1,\ldots,x^n),
\eeq
where
\beq
s(u,x^1,\ldots,x^n)=\frac{1}{T}\; u+ \frac{p}{T}-\sum_{i} \frac{\xi_i}{T}\; x^i,
\eeq
where the intensive variables are now functions of $u,x^1,\ldots,x^n$. 
The Gibbs-Duhem equation becomes
\beq
u\;d\left(\frac{1}{T}\right)+d\left(\frac{p}{T}\right)-\sum_{i}  x^i\; 
d\left(\frac{\xi_i}{T}\right)=0
\eeq
and now only intensive variables appear.\\ 
In the approach by means of the homogeneous 
differential form $\deq$, the use of the densities leads easily 
to the following result:
\beq
\deq=(u+p-\sum_i\; \xi_i\; x^i)\; dV+V\; (du-\sum_i\; \xi_i\; dx^i).
\eeq
Then, we get
\beq
\frac{\deq}{f}=\omega_0+\frac{dV}{V},
\eeq
where 
\beq
\omega_0\equiv \frac{du-\sum_i\; \xi_i\; dx^i}{u+p-\sum_i\; \xi_i\; x^i}=
\frac{ds}{s}. 
\eeq
Notice that in the latter approach both $s$ and $T$ are unknown. 
All the intensive variables $p,\xi_1,\ldots,\xi_n$ are to be known if the 
entropy has to be determined, thus $p$ has to be known in order that 
the fundamental equation can be recovered. This is completely 
in agreement with what happens in the standard approach, where 
only one intensive variable is allowed to be unknown [it can be 
reconstructed by means of the Gibbs-Duhem equation]. 
One may wonder why in the case 
of an homogeneous one-component system one can sometimes recover 
the fundamental equation $S(U,V,N)$ without any knowledge of 
the chemical potential $\mu$. The point is that, 
by passing to the densities with respect to N [indicated again with $u,v$], 
one has $S=N\; s(u,v)$ and 
\beqnl
dS &=& \frac{dU+p\; dV-\mu\; dN}{T} = 
\frac{1}{T(u,v)}\; N \left( du + p(u,v)\; dv \right) +
\frac{1}{T(u,v)}\; \left(u+p(u,v)\; v-\mu (u,v) \right)\; dN\\
&=&
 N\; \frac{du + p(u,v)\; dv}{T(u,v)}\; + 
\frac{u+p(u,v)\; v-\mu (u,v)}{T(u,v)}\; dN\\
&=&
 N\; ds + s\; dN.
\eeqnl
Then, it is possible to find $s(u,v)$ by considering the system 
as closed if the temperature $T(u,v)$ is known. The point is that 
it is legitimate to ignore $\mu$ only if the dependence of the 
temperature on $u,v$ is known.  A typical example of such a system 
where $T$ is known is represented by the classical ideal gas 
\cite{callen}.

\section{conclusions}

In this paper we have shown that, by choosing the extensive 
variables $(U,V,X^1,\ldots,X^n)$ as independent variables, 
the integrability and the homogeneity of $\delta Q_{rev}$ 
allow to find explicitly and immediately an integrating 
factor for $\delta Q_{rev}$. The relation between 
the potential $\has$ and the entropy $S$ is 
straightforward. It is to be noted that the entropy $S$ 
arising from this calculation corresponds to the 
fundamental equation in the entropy representation. 
Thus our approach allows to find a direct link between 
Carath\'eodory approach, based on differential forms, and 
Gibbs one, based on the postulate of the concavity 
and homogeneity of the entropy $S(U,V,X^1,\ldots,X^n)$. 
The role of the homogeneity as a symmetry for the 
system and as a tool for constructing the thermodynamic 
formalism has been enhanced.  

\appendix

\section{integrable homogeneous Pfaffian form}
\label{inhompfaff}

Here we sketch a more elementary proof for some statements 
appearing in sect. I. Let us consider an homogeneous Pfaffian 
form in three variables (the generalization to $n>3$ variables 
is straightforward):
\beq
\omega=P\; dx+Q\; dy+R\; dz,
\label{pfahom}
\eeq
where $P,Q,R$ are homogeneous functions of degree $k$ (non necessarily 
$k$ integer) in $x,y,z$:
\beqnl
&&(x\pa_x\; + y\pa_y\; +z\pa_z)\; P = k\; P;\\
&&(x\pa_x\; + y\pa_y\; +z\pa_z)\; Q = k\; Q;\\
&&(x\pa_x\; + y\pa_y\; +z\pa_z)\; R = k\; R.
\eeqnl
The Liouville vector field is $Y\equiv (x\pa_x\; + y\pa_y\; +z\pa_z)$. 
The integrability 
condition is 
\beq
P\; (\pa_z\; Q-\pa_y\; R)+
Q\; (\pa_x\; R-\pa_z\; P)+
R\; (\pa_y\; P-\pa_x\; Q)=0.
\label{intepfaff}
\eeq
Then an integrating factor for $\omega$ is given by 
\beq
\mu \equiv P\; x+Q\; y+R\; z,
\label{intfhom}
\eeq
provided that $\mu\not \equiv 0$. Let us check that 
\beq
\omega_e\equiv \frac{\omega}{\mu}
\label{exaf}
\eeq
is closed. We have to check the equality of the mixed derivatives, 
that is, we have to verify that 
\beqnl
&&\pa_z\; \left(\frac{P}{\mu}\right)=\pa_x\; \left(\frac{R}{\mu}\right)
\label{verif}\\ 
&&\pa_y\; \left(\frac{P}{\mu}\right)=\pa_x\; \left(\frac{Q}{\mu}\right)\\ 
&&\pa_x\; \left(\frac{Q}{\mu}\right)=\pa_y\; \left(\frac{R}{\mu}\right). 
\eeqnl   
We verify explicitly only (\ref{verif}). We have 
\beqnl
\pa_z\; \left(\frac{P}{\mu}\right)&=&
\frac{1}{\mu}\; \pa_z\; P -\frac{P}{\mu^2}\; 
(x\; \pa_z\; P+y\; \pa_z\; Q+ z\; \pa_z\; R+R)\\
&=&\frac{1}{\mu^2}\; 
(Q\; y\; \pa_z\; P + R\; z\; \pa_z\; P-
P\; y\; \pa_z\; Q- P\; z\; \pa_z\; R+P\; R).
\eeqnl
Analogously, we get
\beq
\pa_x\; \left(\frac{R}{\mu}\right)= 
\frac{1}{\mu^2}\; 
(P\; x\; \pa_x\; R + Q\; y\; \pa_x\; R-
R\; x\; \pa_x\; P- R\; y\; \pa_x\; Q-P\; R).
\eeq 
We have to show that 
\beqnl
&&(Q\; y\; \pa_z\; P + R\; z\; \pa_z\; P-
P\; y\; \pa_z\; Q- P\; z\; \pa_z\; R+P\; R)\cr
&&-
(P\; x\; \pa_x\; R + Q\; y\; \pa_x\; R-
R\; x\; \pa_x\; P- R\; y\; \pa_x\; Q-P\; R)=0.
\eeqnl
We obtain
\beqnl
&&y\; Q\; (\pa_x\; R- \pa_z\; P)
-R\; (x\; \pa_x\; P+z\; \pa_z\; P)
-R\; y\; \pa_y\; Q\cr
&&+P\; (x\; \pa_x\; R+z\; \pa_z\; R)
+P\; y\; \pa_z\; Q=0.
\eeqnl
The homogeneity of $P,R$ allows to modify the second and the 
fourth term above  
\beqnl
&&y\; Q\; (\pa_x\; R- \pa_z\; P)
+R\; y\; \pa_y\; P-k\; P\; R
-R\; y\; \pa_y\; Q\cr
&&-P\; y\; \pa_y\; R+k\; P\; R
+P\; y\; \pa_z\; Q=0.
\eeqnl
This can rewritten as follows
\beq
y\; [
P\; (\pa_z\; Q-\pa_y\; R)+
Q\; (\pa_x\; R-\pa_z\; P)+
R\; (\pa_y\; P-\pa_x\; Q)]=0.
\eeq
Because of the integrability condition (\ref{intepfaff}), 
the term appearing in the square parentheses is zero, 
and the proof is completed. This proof is an extension to the case 
of three variables of the proof appearing in Ref. \cite{ince}, 
p. 19.\\
It is also interesting to note that, an analogous 
generalization of Ref. \cite{ince}, p.18-20, allows to 
show that, if the coefficients $P/\mu,Q/\mu,R/\mu$ of 
the 1-form $\omega_e$ appearing in (\ref{exaf}) 
are homogeneous of degree 
$\alpha\not = -1$, so that $\omega_e$ is an exact 1-form of 
degree $\alpha+1$, 
then the solutions of the equation $\omega_e=0$ 
are 
\beq
g=\hbox{const}.
\eeq
where   
\beq
g\equiv \frac{1}{\alpha+1} \left(
x\; \frac{P}{\mu}+
y\; \frac{Q}{\mu}+
z\; \frac{R}{\mu} \right)
\eeq
is an homogeneous function of degree $\alpha+1$ which 
satisfies 
\beq
\omega_e=dg.
\eeq
In fact, one easily finds that the potential associated 
with $\omega_e$ is given by 
\beq
g=\frac{1}{\alpha+1}\; i_Y (\omega_e)
\label{pothom}
\eeq
[proof: $(\alpha+1)\; dg=d\; i_Y (\omega_e) = 
-i_Y\; d \omega_e + 
L_Y\; \omega_e = L_Y\; \omega_e = (\alpha+1)\; \omega_e$].
If, instead, $\alpha=-1$, then the solutions of $\omega_e=0$ 
have to be found by quadratures \cite{ince}. 

\section{further notes on the integral of $\omega/\mu$}
\label{intomu}

Let us assume that 
\beq
\omega_{(k+1)}\equiv \sum_{i=1}^n\; P_i\; dx^i
\eeq
is an homogeneous 
integrable Pfaffian form of degree $k+1$, with $P_i$ homogeneous 
of degree $k$ for all $i=1,\ldots,n$. The Liouville operator is
\beq
Y=\sum_{i=1}^n\; x^i\; \frac{\pa}{\pa x^i}.
\eeq
If
\beq
\mu=\sum_{i=1}^n\; x^i\; P_i
\label{mufact}
\eeq
is the corresponding integrating factor, 
then 
\beq
d\hat{W}\equiv \frac{\omega_{(k+1)}}{\mu}=\frac{dW}{W}
\label{dimo}
\eeq
where $W>0$ is defined by $\hat{W}=\log(W)$.  
Moreover, 
we know that $W$ has to be an homogeneous function of 
degree $q$, because $d L_Y \hat{W}=L_Y d \hat{W} =0$. 
We then find
\beq
dW = \sum_{i=1}^n\; (\pa_i W)\; dx^i =
\frac{W}{\mu}\; \sum_{i=1}^n\; P_i\; dx^i,
\eeq
thus one has 
\beq
(\pa_i W)= \frac{W}{\mu}\; P_i\quad \forall\; i=1,\ldots,n.
\eeq
As a consequence, one obtains
\beqnl
Y\; W &=& \sum_{i=1}^n\; x^i\; (\pa_i W)\\
&=&\sum_{i=1}^n\; x^i\;  \frac{W}{\mu}\; P_i\\
&=& q\; W.
\eeqnl
From (\ref{mufact}) one finds
\beq
\sum_{i=1}^n\; x^i\;  \frac{W}{\mu}\; P_i = W
\eeq
thus the only allowed value for $q$ is $q=1$. 
If $q=0$, one would get $W=0$ everywhere, which is absurd. 
Then $W$ is homogeneous  of degree one (extensive 
function). One then has $\omega_{(k+1)}=g_{(k)}\; dW$, 
where $g_{(k)}$ is homogeneous of degree $k$. 

We show that, if 
$G$ is an homogeneous function of degree $q$ such that 
\beq
\omega_{(k+1)}=g_{(k+1-q)}\; dG,
\label{omeg}
\eeq
where $g_{(k+1-q)}$ is homogeneous of degree $k+1-q$, 
then necessarily  
\beq
G =\zeta\; W^{q},
\label{homsup}
\eeq
where $\zeta=$ const. In fact, one has
\beq
\omega_{(k+1)}=g_{(k+1-q)}\; dG = g_{(k)}\; dW
\eeq
and 
\beq
g_{(k+1-q)}=\frac{g_{(k)}}{dG/dW}.
\eeq
Moreover, the homogeneity of $G(W)$ implies
\beq
D\; G = q\; G = \frac{dG}{dW}\; W,
\eeq
that is,
\beq
\frac{dG}{dW}=q\; \frac{G}{W},
\eeq
whose solution is (\ref{homsup}). \\ 
Notice that $q=1$ is a consequence of the construction of 
the integrating factor $\mu$. In fact, one could find another 
integrating factor ${\bar {\mu}}=\mu/q$. The corresponding 
potential would be then $G=W^q$.

\section{implementation of assumption h8)}
\label{hyph8}

Let us assume that assumption $\overline{{\mathrm h7})}$ holds and use the 
variable $B\equiv U-b(V,X^1,\ldots,X^n)$. 
We then have ${\cal D}=(0,\infty)\times {\cal K}_b$. 
We further assume that the third law (\ref{planck}) holds and that 
$\lim_{U\to \infty}\; S(U,V,X^1,\ldots,X^n)=+\infty$ for all fixed 
$(V,X^1,\ldots,X^n)\in {\cal K}_b$. The latter hypothesis is 
reasonable [notice that it does not hold for systems which allow 
negative temperatures]. Then, 
for each positive constant $c\in \RR_+$, one finds for all fixed 
$(V,X^1,\ldots,X^n)\in {\cal K}_b$ 
\beqnl
&&\lim_{B\to 0^+}\ (S(B,V,X^1,\ldots,X^n)-c)=-c<0\\
&&\cr
&&\lim_{B\to +\infty} (S(B,V,X^1,\ldots,X^n)-c)=+\infty.
\eeqnl
Then, a variant of the implicit function theorem \cite{roux} shows that 
there exists a unique function\\ 
$B_c (V,X^1,\ldots,X^n):{\cal K}_b\to (0,\infty)$ such that 
$S(B_c (V,X^1,\ldots,X^n),V,X^1,\ldots,X^n)-c=0$.\\ 
This function 
is continuous (actually $C^2$) and it defines 
the unique leaf corresponding to $S^{-1}(c)$. 

[Sketch of proof: Let us put $Y\equiv V,X^1,\ldots,X^n$ in 
the following. Then $Y\in {\cal K}_b$. 
$S$ is continuous and monotonically strictly increasing in $B$, hence  
also the function $\sigma_c (B,Y)\equiv S(B,Y) -c$ has these properties.  
In particular, $\sigma_c (B,Y)$ is continuous as 
a function of $B$ for any fixed $Y\in {\cal K}_b$.  
Permanence of the sign implies that there exist 
two real numbers $b_1<b_2$ such that 
$\sigma_c (b_1,Y)<0< \sigma_c (b_2,Y)$ for 
all $Y\in {\cal K}_b$. Then, from 
the intermediate value theorem it follows that there exists a 
value $\bar{b}\in (b_1,b_2)$ such that 
$\sigma_c (\bar{b},Y)=0$ for any fixed 
$Y\in {\cal K}_b$.  Monotonicity 
ensures that $\bar{b}$ is unique. The function $B_c (Y)$ is then 
defined as the map $B_c (Y)=\bar{b}$ 
for each fixed $Y\in {\cal K}_b$. 
This function is also continuous. In fact, the set 
$Z(\sigma_c)\equiv \{(B,Y)|\sigma_c (B,Y)=0 \}$ is a closed 
set, being $\sigma_c$ a continuous function. 
Given a sequence $\{ Y_n \}\subseteq {\cal K}_b$ such that 
$\lim_{n\to \infty} Y_n=Y_0\in {\cal K}_b$, one has 
$(B_c (Y_n),Y_n)\in Z(\sigma_c)$ and, moreover, 
$(\lim_{n\to \infty} B_c (Y_n),\lim_{n\to \infty} Y_n)\in Z(\sigma_c)$, 
being $Z(\sigma_c)$ closed. 
$\lim_{n\to \infty} B_c (Y_n)=B_c (Y_0)$ then follows.]

\end{document}